\documentclass[submission,copyright,creativecommons]{eptcs}
\providecommand{\event}{TTC 2011} 

\usepackage{multirow}
\usepackage{graphicx}
\usepackage{subfigure}
\usepackage{url}
\usepackage{amssymb}
\usepackage{xspace}

\usepackage{amsmath}
\usepackage{listings}

\makeindex

\newcommand{\myparagraph}[1]{\vspace{.5em}\par\noindent\textbf{#1}}

\lstdefinelanguage{groovy} {
    emph={println, new, tokenize, each, def, static, for, if, else, in, assert},
    emphstyle=\bfseries,
    morecomment=[l]{//},
    basicstyle=\fontfamily{pcr}\scriptsize,
    string=[b]",
    showstringspaces=false
}

\newcommand{\operation}[1]{\textsf{#1}}

\title{Saying Hello World with Edapt - A Solution to the TTC 2011 Instructive Case}
\author{Markus Herrmannsdoerfer
\institute{Institut f\"ur Informatik,
  Technische Universit\"{a}t M\"{u}nchen 
}
\email{herrmama@in.tum.de}
}
\def\titlerunning{Saying Hello World with Edapt - A Solution to the TTC 2011 Instructive Case}
\def\authorrunning{M. Herrmannsdoerfer}
\begin{document}
\maketitle

\begin{abstract}
This paper gives an overview of the Edapt solution to the hello world case \cite{helloworldcase} of the Transformation Tool Contest 2011.
\end{abstract}

\section{Edapt in a Nutshell}

Edapt\footnote{\url{http://www.eclipse.org/edapt}} is a transformation tool tailored for the migration of models in response to metamodel adaptation.
Edapt is an official Eclipse tool derived from the research prototype COPE.

\begin{figure}[htb]
\centering
		\includegraphics[scale=0.6]{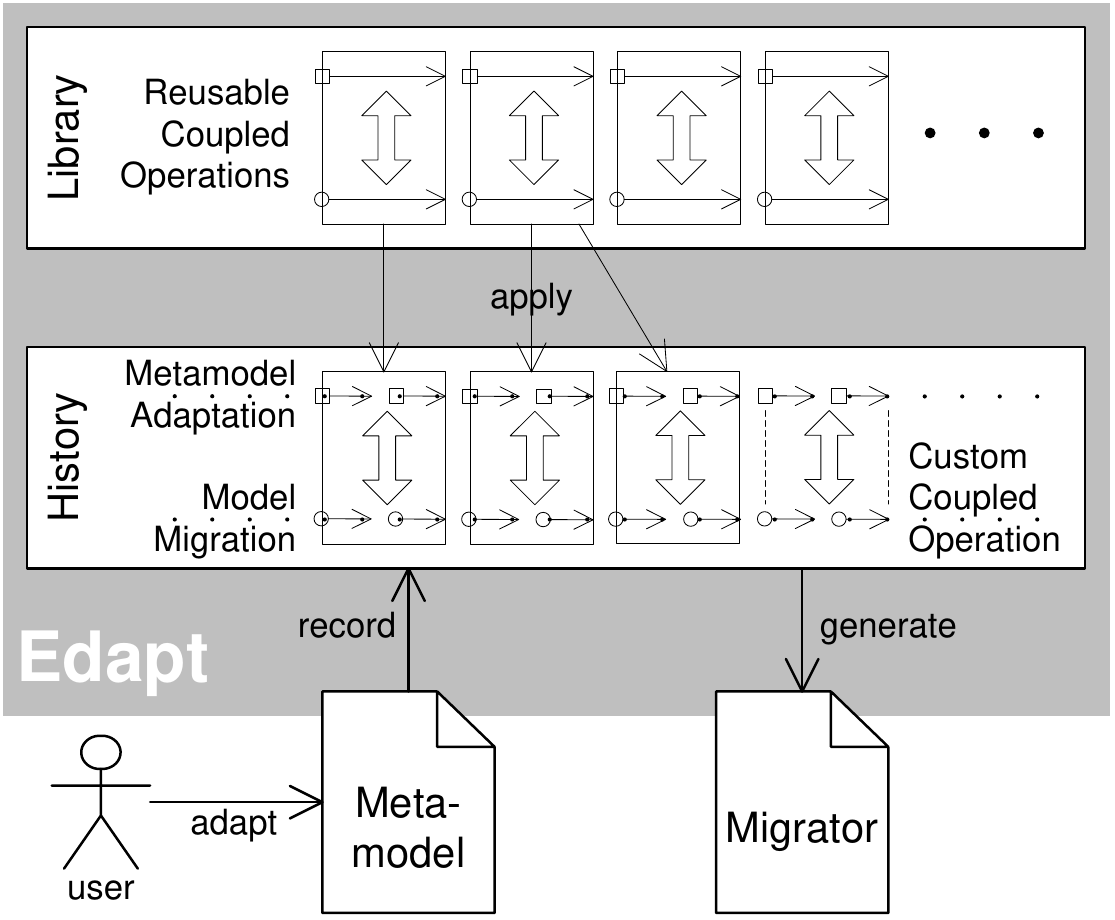}
				\vskip -5pt
\caption{Overview of Edapt}
\label{fig:overview}
\end{figure}

\myparagraph{Modeling the Coupled Evolution.}
As depicted by Figure~\ref{fig:overview}, Edapt specifies the me\-ta\-mo\-del adaptation as a sequence of operations in an explicit history model.
The operations can be enriched with instructions for model migration to form so-called coupled operations.
Edapt provides two kinds of coupled operations according to the automatability of the model migration~\cite{Herrmannsdoerfer2009_COPE-AutomatingCoupledEvolutionofMetamodelsandModels}: reusable and custom coupled operations.

Reuse of recurring migration specifications allows to reduce the effort associated with building a model migration~\cite{Herrmannsdoerfer2008_AutomatabilityofCoupledEvolutionofMetamodelsandModelsinPractice}.
Edapt thus provides \emph{reusable coupled operations} which make metamodel adaptation and model migration independent of the specific metamodel through parameters and constraints restricting the applicability of the operation.
An example for a reusable coupled operation is \emph{Enumeration to Sub Classes} which replaces an enumeration attribute with subclasses for each literal of the enumeration.
Currently, Edapt comes with a library of over 60 reusable coupled operations~\cite{Herrmannsdoerfer2010_AnExtensiveCatalogofOperatorsfortheCoupledEvolutionofMetamodelsandModels}.
By means of studying real-life metamodel histories, we have shown that, in practice, most of the coupled evolution can be covered by reusable coupled operations~\cite{Herrmannsdoerfer2008_AutomatabilityofCoupledEvolutionofMetamodelsandModelsinPractice,Herrmannsdoerfer2010_LanguageEvolutioninPracticeTheHistoryofGMF}.

Migration specifications can become so specific to a certain metamodel that reuse does not make sen\-se~\cite{Herrmannsdoerfer2008_AutomatabilityofCoupledEvolutionofMetamodelsandModelsinPractice}.
To express these complex migrations, Edapt allows the user to define a custom coupled operation by manually encoding a model migration for a metamodel adaptation in a Turing-complete language~\cite{Herrmannsdoerfer2008_COPEALanguagefortheCoupledEvolutionofMetamodelsandModels}.
By softening the conformance of the model to the metamodel within a coupled operation, both metamodel adaptation and model migration can be specified as in-place transformations, requiring only to specify the difference.
A transaction mechanism ensures conformance at the boundaries of the coupled operation.

\begin{figure}[!tb]
\centering
		\includegraphics[width=.85\textwidth]{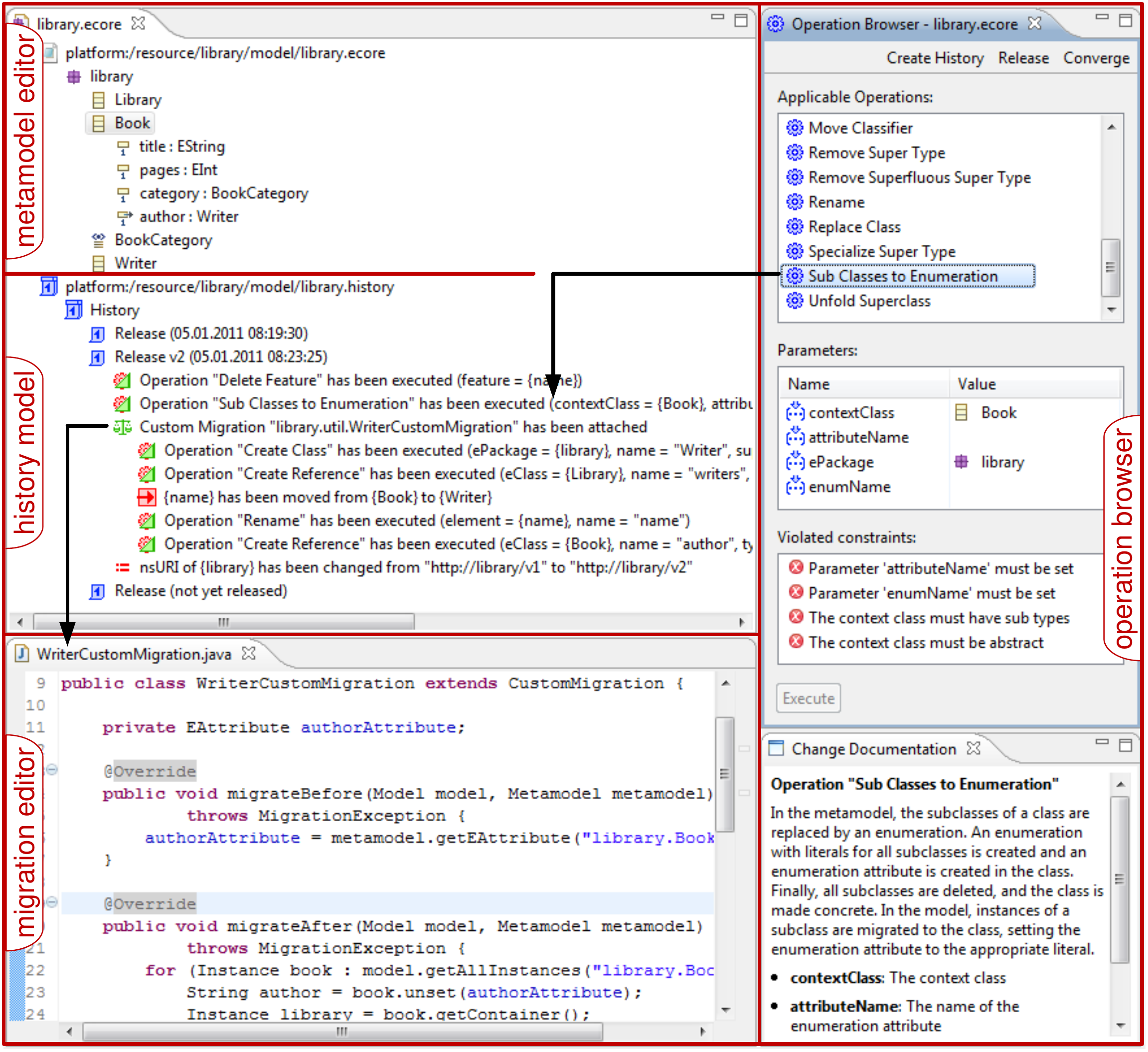}
		\vskip -5pt
\caption{User interface of Edapt}
\label{fig:screenshot}
\end{figure}

\myparagraph{Recording the Coupled Evolution.}
To not lose the intention behind the metamodel adaptation, Edapt is intended to be used already when adapting the metamodel.
Therefore, Edapt's user interface, which is depicted in Figure~\ref{fig:screenshot}, is directly integrated into the existing EMF \emph{metamodel editor}.
The user interface provides access to the \emph{history model} in which Edapt records the sequence of coupled operations.
An initial history can be created for an existing metamodel by invoking \emph{Create History} in the \emph{operation browser} which also allows the user to \emph{Release} the metamodel.

The user can adapt the metamodel by applying reusable coupled operations through the \emph{operation browser}.
The operation browser allows to set the parameters of a reusable coupled operation, and gives feedback on the operation's applicability based on the constraints.
When a reusable coupled operation is executed, its application is automatically recorded in the history model.
Figure~\ref{fig:screenshot} shows the operation \operation{Sub Classes to Enumeration} being selected in the operation browser and recorded to the history model.

The user needs to perform a custom coupled operation only, in case no reusable coupled operation is available for the change at hand.
First, the metamodel is directly adapted in the metamodel editor, in response to which the changes are automatically recorded in the history.
A migration can later be attached to the sequence of metamodel changes.
Figure~\ref{fig:screenshot} shows the \emph{migration editor} to encode the custom migration in Java.

\section{Hello World Case}


Since Edapt is tailored for model migration, the migration tasks could be solved using only reusable coupled operations.
For all other tasks, custom coupled operations are required, as Edapt is not tailored for these cases.

\begin{figure}[tb]
	\centering
		\includegraphics[scale=0.6]{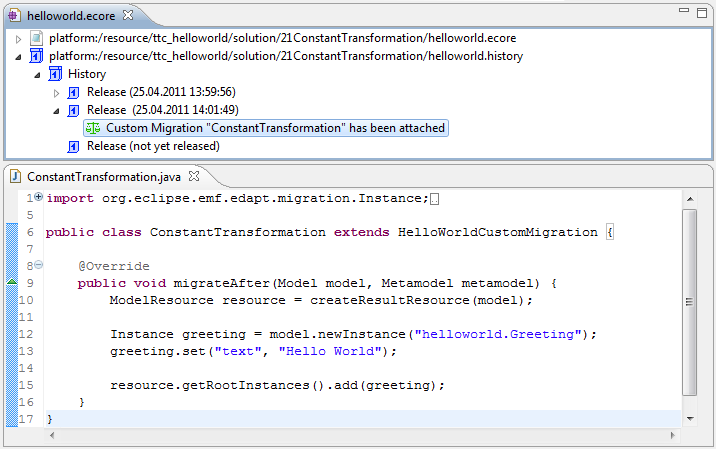}
		\vskip -5pt
		\caption{History model with a custom migration}
	\label{fig:custommigration}
\end{figure}

Figure~\ref{fig:custommigration} shows how the history model looks like for all tasks of this case that are solved using custom coupled operations.
In this case, the custom coupled operation always consists of a custom migration which is attached to an empty metamodel adaptation.
The custom migration is implemented as a Java class that inherits from a special super class.

The complete solutions are available in the appendix, through a SHARE demo \cite{helloworldsolutionedaptshare} and in the repository of the Eclipse Edapt project\footnote{\url{http://dev.eclipse.org/svnroot/modeling/org.eclipse.emft.edapt/trunk/examples/ttc_helloworld}}.
In the following, we explain the main characteristics of the solutions for the different tasks.

\myparagraph{2.1 Hello World!}
Figure~\ref{fig:custommigration} shows how the constant transformation is implemented using the migration language provided by Edapt.
Since Edapt is a migration tool, the transformation is always performed in-place.
To store the result at another location, we use helper methods that are provided by the superclass \textsf{HelloWorldCustomMigration}.
The task to perform the extended constant transformation is solved in a similar way.
For the model-to-text-transformation, we also have to include the result metamodel in the history and provide helper methods to store instances of the classes defined by this metamodel.

\myparagraph{2.2 Count Matches with certain Properties.}
All the count tasks require the result metamodel to be part of the history and a helper method to store the integer result which is provided by the superclass \textsf{HelloWorldCustomMigration}.
The solutions of the tasks to count the number of nodes, looping edges, isolated nodes and dangling edges are straightforward.
For the solution of the task to count the number of circles, we implemented the helper method \textsf{getReachable} to get the nodes reachable from a node through directed edges.
For this helper method, we used the function \textsf{getInverse} to navigate the inverse of an association.

\myparagraph{2.3 Reverse Edges.}
The solution to this task is straightforward, since we only have to exchange \textsf{src} and \textsf{trg} of each \textsf{Edge}.

\begin{figure}[tb]
	\centering
		\includegraphics[scale=0.6]{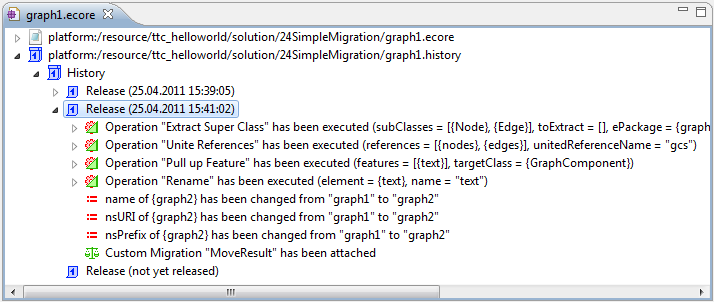}
		\vskip -5pt
		\caption{History model for the simple migration}
	\label{fig:simplemigration}
\end{figure}

\myparagraph{2.4 Simple Migration.}
Figure~\ref{fig:simplemigration} shows the history model for the simple migration which can be solved completely using already available reusable coupled operations.
Note that the custom coupled operation is only necessary to store the result of a transformation in a different file.
First, the common super class \textsf{GraphComponent} is created for classes \textsf{Node} and \textsf{Edge}.
Then, the associations \textsf{nodes} and \textsf{edges} are united into the association \textsf{gcs}.
Finally, the attribute \textsf{name} is pulled up from class \textsf{Node} to \textsf{GraphComponent} and renamed to \textsf{text}.

\begin{figure}[tb]
	\centering
		\includegraphics[scale=0.6]{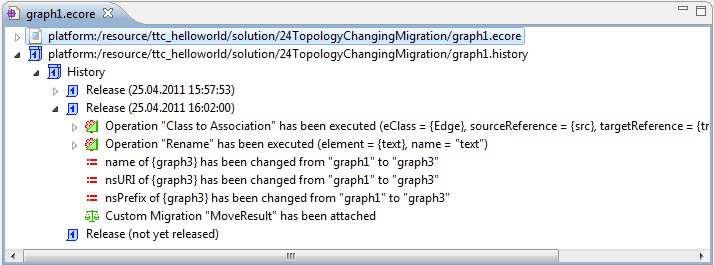}
		\vskip -5pt
		\caption{History model for the topology-changing migration}
	\label{fig:topologychangingmigration}
\end{figure}

Figure~\ref{fig:topologychangingmigration} shows the history model for the topology-changing migration which can also be solved using reusable coupled operations.
The operation \textsf{ClassToAssociation} is applied to replace the class \textsf{Edge} by the association \textsf{linksTo}.
Finally, also a \textsf{Rename} of an attribute is required to complete the migration.

\myparagraph{2.5 Delete Node with Specific Name and its Incident Edges.}
This task can also be implemented quite easily, since Edapt provides a method to \textsf{delete} instances of classes.
To also delete all incident edges, we can again use the method \textsf{getInverse} to navigate to the edges which have the node as source or target.

\myparagraph{2.6 Insert Transitive Edges.}
The solution to this task is a little bit more involved.
To not let the newly created edges influence the result, we first determine the pairs of nodes for which edges need to be created.
Here, we can again rely on our helper method \textsf{getReachable} to obtain the nodes reachable from a node.
Finally, we create the edges for these nodes.

\section{Conclusion}

As Edapt is a transformation tool targeted at model migration, it clearly shows its strengths in the migration tasks.
The migration tasks can be solved by applying only reusable coupled operations.
Thereby, not a single line of custom migration code needs to be written.

Although a degenerated case, the other tasks can be solved by attaching custom migrations to an empty metamodel adaptation.
The custom migrations are implemented in Java based on the API provided by Edapt to navigate and modify models.
Even though the Java solutions are quite concise and clear, a specialized DSL could further improve conciseness and clarity.
However, we can rely on Java's abstraction mechanisms to organize the implementation, and on the strong Java tooling to implement, refactor and debug the solution.


\bibliographystyle{eptcs}
\bibliography{literature}

\appendix

\section{Solution}

\subsection*{2.1-2.6 Base Class for Custom Migrations for the Hello World Case}
\begin{lstlisting}[language=java]
import org.eclipse.emf.common.util.URI;
import org.eclipse.emf.edapt.migration.CustomMigration;
import org.eclipse.emf.edapt.migration.Instance;
import org.eclipse.emf.edapt.migration.Model;
import org.eclipse.emf.edapt.migration.ModelResource;

public abstract class HelloWorldCustomMigration extends CustomMigration {

	/** Create the resource in which to store the result of the transformation. */
	protected ModelResource createResultResource(Model model) {
		URI resultUri = getResultURI(model);
		return model.newResource(resultUri);
	}

	/** Change the location in which the model is stored to the result location. */
	protected void moveResult(Model model) {
		model.getResources().get(0).setUri(getResultURI(model));
	}

	/** Get the location in which the result should be stored. */
	private URI getResultURI(Model model) {
		URI uri = model.getResources().get(0).getUri();
		URI resultUri = uri
				.trimSegments(1)
				.appendSegment(uri.trimFileExtension().lastSegment() + "result")
				.appendFileExtension(uri.fileExtension());
		return resultUri;
	}

	/** Create the result resource and save a result of type integer in it. */
	protected void saveResult(Model model, int i) {
		ModelResource resource = createResultResource(model);
		Instance instance = model.newInstance("result.IntResult");
		instance.set("result", i);
		resource.getRootInstances().add(instance);
	}

	/** Save a result of type String in a resource. */
	protected void saveResult(ModelResource resource, String s) {
		Instance instance = resource.getModel().newInstance(
				"result.StringResult");
		instance.set("result", s);
		resource.getRootInstances().add(instance);
	}
}
\end{lstlisting}
\vspace{1em}

\subsection*{2.1 Constant Transformation}
\begin{lstlisting}[language=java]
import org.eclipse.emf.edapt.migration.Instance;
import org.eclipse.emf.edapt.migration.Metamodel;
import org.eclipse.emf.edapt.migration.Model;
import org.eclipse.emf.edapt.migration.ModelResource;

public class ConstantTransformation extends HelloWorldCustomMigration {

	@Override
	public void migrateAfter(Model model, Metamodel metamodel) {
		ModelResource resource = createResultResource(model);

		Instance greeting = model.newInstance("helloworld.Greeting");
		greeting.set("text", "Hello World");

		resource.getRootInstances().add(greeting);
	}
}
\end{lstlisting}
\vspace{1em}

\subsection*{2.1 Constant Transformation to create Model with References}
\begin{lstlisting}[language=java]
import org.eclipse.emf.edapt.migration.Instance;
import org.eclipse.emf.edapt.migration.Metamodel;
import org.eclipse.emf.edapt.migration.Model;
import org.eclipse.emf.edapt.migration.ModelResource;

public class ConstantTransformationReferences extends HelloWorldCustomMigration {

	@Override
	public void migrateAfter(Model model, Metamodel metamodel) {
		ModelResource resource = createResultResource(model);
		
		metamodel.setDefaultPackage("helloworldext");

		Instance greeting = model.newInstance("Greeting");

		Instance message = model.newInstance("GreetingMessage");
		message.set("text", "Hello");
		greeting.set("greetingMessage", message);

		Instance person = model.newInstance("Person");
		greeting.set("person", person);
		person.set("name", "TTC Participants");

		resource.getRootInstances().add(greeting);
	}
}
\end{lstlisting}
\vspace{1em}

\subsection*{2.1 Model-to-Text-Transformation}
\begin{lstlisting}[language=java]
import org.eclipse.emf.edapt.migration.Instance;
import org.eclipse.emf.edapt.migration.Metamodel;
import org.eclipse.emf.edapt.migration.Model;
import org.eclipse.emf.edapt.migration.ModelResource;

public class ModelToTextTransformation extends HelloWorldCustomMigration {

	@Override
	public void migrateBefore(Model model, Metamodel metamodel) {
		ModelResource resource = createResultResource(model);
		metamodel.setDefaultPackage("helloworldext");
		for (Instance greeting : model.getAllInstances("Greeting")) {
			String greetingText = greeting.getLink("greetingMessage").get(
					"text");
			Object personName = greeting.getLink("person").get("name");
			String text = greetingText + " " + personName + "!";
			saveResult(resource, text);
		}
	}
}
\end{lstlisting}
\vspace{1em}

\subsection*{2.2 Base Class for Custom Migrations for the Graph1 Metamodel}
\begin{lstlisting}[language=java]
import java.util.ArrayList;
import java.util.List;

import org.eclipse.emf.edapt.migration.Instance;

public abstract class Graph1CustomMigration extends HelloWorldCustomMigration {

	/** Get the nodes reachable from a nodes through directed edges. */
	protected List<Instance> getReachable(Instance node) {
		List<Instance> reachable = new ArrayList<Instance>();
		for (Instance edge : node.getInverse("graph1.Edge.src")) {
			reachable.add(edge.getLink("trg"));
		}
		return reachable;
	}
}
\end{lstlisting}
\vspace{1em}

\subsection*{2.2 Count the Number of Nodes}
\begin{lstlisting}[language=java]
import org.eclipse.emf.edapt.migration.Metamodel;
import org.eclipse.emf.edapt.migration.Model;

public class CountNodes extends HelloWorldCustomMigration {

	@Override
	public void migrateBefore(Model model, Metamodel metamodel) {
		int nodes = model.getAllInstances("graph1.Node").size();
		saveResult(model, nodes);
	}
}
\end{lstlisting}
\vspace{1em}

\subsection*{2.2 Count the number of looping Edges}
\begin{lstlisting}[language=java]
import org.eclipse.emf.edapt.migration.Instance;
import org.eclipse.emf.edapt.migration.Metamodel;
import org.eclipse.emf.edapt.migration.Model;

public class CountLoopingEdges extends HelloWorldCustomMigration {

	@Override
	public void migrateAfter(Model model, Metamodel metamodel) {
		int loops = 0;
		for (Instance edge : model.getAllInstances("graph1.Edge")) {
			if (edge.get("src") == edge.get("trg")) {
				loops++;
			}
		}
		saveResult(model, loops);
	}
}
\end{lstlisting}
\vspace{1em}

\subsection*{2.2 Count the number of isolated Nodes}
\begin{lstlisting}[language=java]
import org.eclipse.emf.edapt.migration.Instance;
import org.eclipse.emf.edapt.migration.Metamodel;
import org.eclipse.emf.edapt.migration.Model;

public class CountIsolatedNodes extends HelloWorldCustomMigration {

	@Override
	public void migrateBefore(Model model, Metamodel metamodel) {
		metamodel.setDefaultPackage("graph1");
		int isolated = 0;
		for (Instance node : model.getAllInstances("Node")) {
			if (node.getInverse("Edge.src").isEmpty()
					&& node.getInverse("Edge.trg").isEmpty()) {
				isolated++;
			}
		}
		saveResult(model, isolated);
	}
}
\end{lstlisting}
\vspace{1em}

\subsection*{2.2 Count the Number of Circles consisting of three Nodes}
\begin{lstlisting}[language=java]
import org.eclipse.emf.edapt.migration.Instance;
import org.eclipse.emf.edapt.migration.Metamodel;
import org.eclipse.emf.edapt.migration.Model;

public class CountCircles extends Graph1CustomMigration {

	@Override
	public void migrateBefore(Model model, Metamodel metamodel) {
		metamodel.setDefaultPackage("graph1");
		int circles = 0;
		for (Instance n1 : model.getAllInstances("Node")) {
			for (Instance n2 : getReachable(n1)) {
				if (n1 != n2) {
					for (Instance n3 : getReachable(n2)) {
						if (n2 != n3 && n1 != n3) {
							if (getReachable(n3).contains(n1)) {
								circles++;
							}
						}
					}
				}
			}
		}
		saveResult(model, circles);
	}
}
\end{lstlisting}
\vspace{1em}

\subsection*{2.2 Count the Number of dangling Edges}
\begin{lstlisting}[language=java]
import org.eclipse.emf.edapt.migration.Instance;
import org.eclipse.emf.edapt.migration.Metamodel;
import org.eclipse.emf.edapt.migration.Model;

public class CountDanglingEdges extends HelloWorldCustomMigration {

	@Override
	public void migrateBefore(Model model, Metamodel metamodel) {
		int dangling = 0;
		for (Instance edge : model.getAllInstances("graph1.Edge")) {
			if (edge.get("src") == null || edge.get("trg") == null) {
				dangling++;
			}
		}
		saveResult(model, dangling);
	}
}
\end{lstlisting}
\vspace{1em}

\subsection*{2.3 Reverse Edges}
\begin{lstlisting}[language=java]
import org.eclipse.emf.edapt.migration.Instance;
import org.eclipse.emf.edapt.migration.Metamodel;
import org.eclipse.emf.edapt.migration.Model;

public class ReverseEdges extends HelloWorldCustomMigration {

	@Override
	public void migrateBefore(Model model, Metamodel metamodel) {
		moveResult(model);
		
		for (Instance edge : model.getAllInstances("graph1.Edge")) {
			Instance src = edge.get("src");
			Instance trg = edge.get("trg");
			edge.set("src", trg);
			edge.set("trg", src);
		}
	}
}
\end{lstlisting}
\vspace{1em}

\subsection*{2.4 Custom Migration to Move the Result}
\begin{lstlisting}[language=java]
import org.eclipse.emf.edapt.migration.Metamodel;
import org.eclipse.emf.edapt.migration.MigrationException;
import org.eclipse.emf.edapt.migration.Model;


public class MoveResult extends HelloWorldCustomMigration {

	@Override
	public void migrateBefore(Model model, Metamodel metamodel)
			throws MigrationException {
		moveResult(model);
	}
}
\end{lstlisting}
\vspace{1em}

\subsection*{2.4 Simple Migration}
see Figure~\ref{fig:simplemigration}

\subsection*{2.4 Topology-Changing Migration}
see Figure~\ref{fig:topologychangingmigration}

\subsection*{2.5 Delete Node with name and incident Edges}
\begin{lstlisting}[language=java]
import org.eclipse.emf.edapt.migration.Instance;
import org.eclipse.emf.edapt.migration.Metamodel;
import org.eclipse.emf.edapt.migration.MigrationException;
import org.eclipse.emf.edapt.migration.Model;

public class DeleteNodeWithName extends HelloWorldCustomMigration {

	@Override
	public void migrateBefore(Model model, Metamodel metamodel)
			throws MigrationException {
		moveResult(model);

		metamodel.setDefaultPackage("graph1");
		for (Instance node : model.getAllInstances("Node")) {
			if ("n1".equals(node.get("name"))) {
				for (Instance edge : node.getInverse("Edge.src")) {
					model.delete(edge);
				}
				for (Instance edge : node.getInverse("Edge.trg")) {
					model.delete(edge);
				}
				model.delete(node);
			}
		}
	}
}
\end{lstlisting}
\vspace{1em}

\subsection*{2.6 Insert Transitive Edges}
\begin{lstlisting}[language=java]
import java.util.ArrayList;
import java.util.Arrays;
import java.util.List;

import org.eclipse.emf.edapt.migration.Instance;
import org.eclipse.emf.edapt.migration.Metamodel;
import org.eclipse.emf.edapt.migration.Model;

public class InsertTransitiveEdges extends Graph1CustomMigration {

	@Override
	public void migrateBefore(Model model, Metamodel metamodel) {
		moveResult(model);

		metamodel.setDefaultPackage("graph1");
		List<List<Instance>> pairs = new ArrayList<List<Instance>>();
		for (Instance n1 : model.getAllInstances("Node")) {
			for (Instance n2 : getReachable(n1)) {
				for (Instance n3 : getReachable(n2)) {
					pairs.add(Arrays.asList(n1, n3));
				}
			}
		}
		Instance graph = model.getAllInstances("Graph").get(0);
		for (List<Instance> pair : pairs) {
			Instance n1 = pair.get(0);
			Instance n3 = pair.get(1);
			if (!getReachable(n1).contains(n3)) {
				Instance edge = model.newInstance("Edge");
				edge.set("src", n1);
				edge.set("trg", n3);
				graph.add("edges", edge);
			}
		}
	}
}
\end{lstlisting}

\end{document}